\UseRawInputEncoding
\documentclass[12pt]{spieman}  
\usepackage{amsmath,amsfonts,amssymb}
\usepackage{threeparttable}
\usepackage{graphicx}
\usepackage{setspace}
\usepackage{tocloft}
\usepackage{xcolor}
\usepackage[numbers]{natbib}
\usepackage{aas_macros}
\usepackage{colortbl}
\usepackage {lineno}

\newcommand{\hi}{\ifmmode\mathrm{H\,{\scriptscriptstyle I}}\else{}H\,{\scriptsize I}\fi}

\title{SKA-Low Intensity Mapping Pathfinder Updates: Deeper 21\,cm Power Spectrum Limits from Improved Analysis Frameworks}

\author[a,b]{Nichole Barry}
\author[c,d,e]{Gianni Bernardi}
\author[a,b]{Bradley Greig}
\author[f*]{Nicholas Kern}
\author[g,h]{Florent Mertens}
\affil[a]{The University of Melbourne, School of Physics, Parkville, VIC 3010, Australia}
\affil[b]{ARC Centre of Excellence for All Sky Astrophysics in 3 Dimensions (ASTRO-3D)}
\affil[c]{INAF - Istituto di Radioastronomia, via Gobetti 101, 40129 Bologna, Italy}
\affil[d]{Department of Physics and Electronics, Rhodes University, PO Box 94, Grahamstown, 6140, South Africa}
\affil[e]{South African Radio Astronomy Observatory, Black River Park, 2 Fir Street, Observatory, Cape Town, 7925, South Africa}
\affil[f]{Dept. of Physics and Kavli Institute for Astrophysics and Space Research, Massuchusetts Institute of Technology, Cambridge, MA, USA}
\affil[g]{LERMA, Observatoire de Paris, PSL Research University, CNRS, Sorbonne Universit\'e, F-75014 Paris, France}
\affil[h]{Kapteyn Astronomical Institute, University of Groningen, PO Box 800, 9700 AV Groningen, The Netherlands}

\cftpagenumbersoff{figure}
\cftpagenumbersoff{table} 
\begin{document} 
\maketitle

\begin{abstract}

The Square Kilometre Array (SKA) is a planned radio interferometer of unprecedented scale that will revolutionize low-frequency radio astronomy when completed. In particular, one of its core science drivers is the systematic mapping of the Cosmic Dawn and Epoch of Reionization, which mark the birth of the first stars and galaxies in the Universe and their subsequent ionization of primordial intergalactic hydrogen, respectively. The SKA will offer the most sensitive view of these poorly understood epochs using the redshifted 21\,cm hyperfine signal from intergalactic hydrogen. However, significant technical challenges stand in the way of realizing this scientific promise. 
These mainly involve the mitigation of systematics coming from astrophysical foregrounds, terrestrial radio interference, and the instrumental response. The Low Frequency Array, the Murchison Widefield Array and the Hydrogen Epoch of Reionization Array are SKA pathfinder experiments that have developed a variety of strategies for addressing these challenges, each with unique characteristics that stem largely from their different instrumental designs. We outline these various directions, highlighting key differences and synergies, and discuss how these relate to the future of low-frequency intensity mapping with the SKA. We also briefly summarize the challenges associated with modeling the 21\,cm signal and discuss the methodologies being proposed for inferring constraints on astrophysical models.

\end{abstract}


{\noindent \footnotesize\textbf{*}Nicholas Kern,  \linkable{nkern@mit.edu}}

\begin{spacing}{2}   

\section{Introduction}
\label{sect:intro}

Low frequency radio observations ($50 < \nu < 200$ MHz) of the redshifted 21\,cm line from neutral Hydrogen are some of the most promising probes of the Epoch of Reionization (EoR) and the era marking the birth of the first luminous structures in the Universe, known as Cosmic Dawn (CD) \citep{Ciardi2005,Furlanetto2006,Morales2010,Pritchard2012}.
The aim of these radio observations, known as intensity mapping surveys, is to scan the sky to produce maps of 21\,cm emission, and to use their large frequency bandwidths as a tool to probe the evolution of the 21\,cm signal across redshift.
These measurements would transform our understanding of the properties of the first stars and galaxies, and shed light on the evolution of primordial gas in the intergalactic medium (IGM).
In addition, 21\,cm observations of the EoR and CD can be used in conjunction with other cosmological probes to help pin down constraints on $\Lambda$CDM cosmology \citep{Liu2020}.

However, a key technical challenge plagues these efforts; contaminating galactic and extra-galactic foreground emission is orders of magnitude brighter than the underlying cosmological signal of interest.
This sets up a delicate signal separation problem, which is compounded by the fact that neither the 21\,cm cosmological signal, the low-frequency foreground sky, nor the instrumental response of the telescope are known a priori by the observer to high precision.
This forms the basis of the 21\,cm foreground separation problem.
Although the early view of the problem was to model the spatial and spectral properties of the foregrounds in order to subtract them  \citep[e.g.,][]{Morales2006,Bowman2009,Bernardi2010,Chapman2013}, the last decade has seen a shift towards an ``avoidance vs subtraction" paradigm: the former approach assumes foregrounds cannot be modeled and simply discards contaminated data, whereas the latter attempts to carefully clean the data of contamination through precise sky modeling.

Foreground emission is generally smooth as a function of frequency owing to its largely non-thermal origin, whereas the 21\,cm signal is highly spectrally variant, as it probes the inhomogeneities of the IGM along the observer's line-of-sight.
Because of this, foregrounds occupy a well defined, wedge-like region in the transverse and line-of-sight Fourier domains ($k_\perp - k_\parallel$, respectively),\footnote{$k_\perp$ is proportional to baseline length and $k_\parallel$ is proportional to the Fourier conjugate of frequency \citep{morales_toward_2004}.} leaving an uncontaminated region where the high-redshift 21\,cm signal is theoretically free of contaminants \citep[the ``EoR window";][]{Datta:2012,Parsons2012,Morales2012,Pober2013}. This ideal scenario is, in practice, degraded by any instrumental response that corrupts the smooth-spectrum foregrounds, eventually leaking its power into the otherwise pristine EoR window. The foreground separation problem then turns into a joint foreground characterization and instrument calibration problem, i.e., how to best correct the instrumental response and simultaneously preserve the foreground spectral smoothness while modeling and improving our understanding of the low-frequency sky.

\begin{figure*}
\centering
\includegraphics[width=\linewidth]{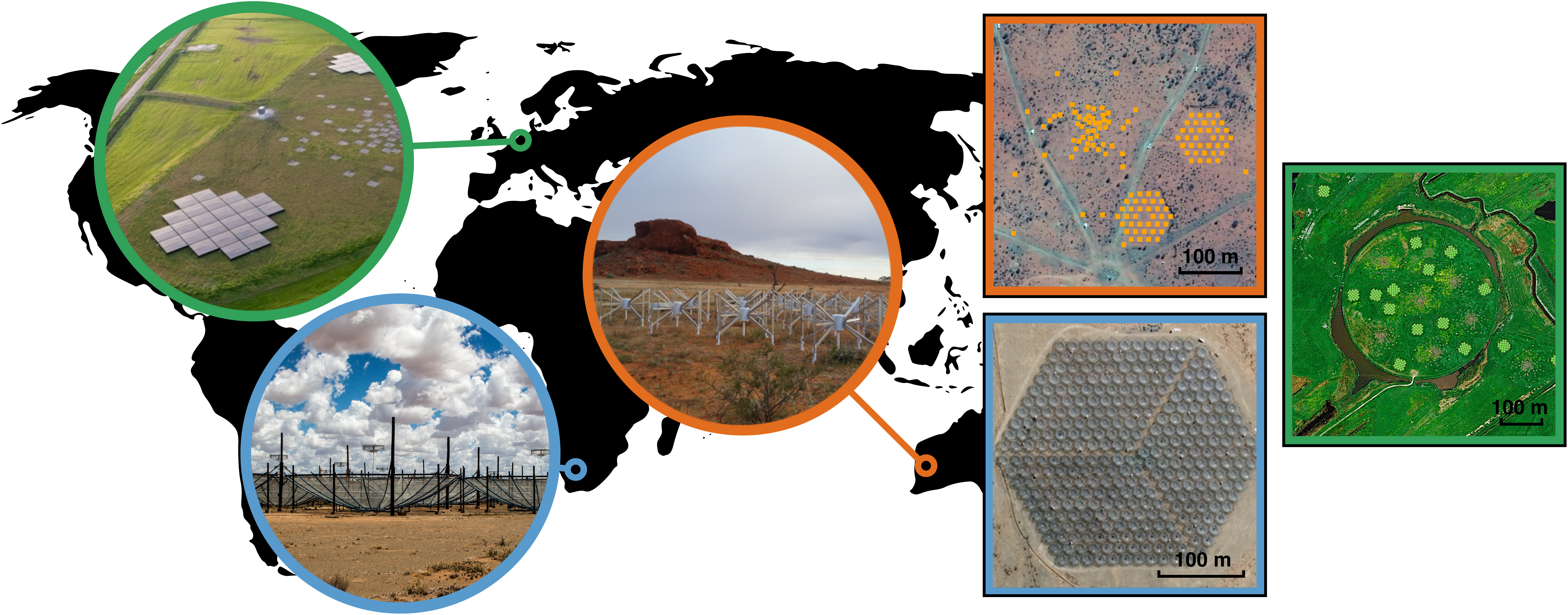}
\caption{The three SKA-Low intensity mapping pathfinder experiments, including (from left to right) LOFAR, HERA, and the MWA (Phase II EoR layout). We show the array layouts of the experiments (right insets), demonstrating the differences in their antenna array layout. Satellite image credit: Google and Maxar Technologies.}
\label{fig:telescopes}
\end{figure*}


As the next generation of EoR and CD intensity mapping telescopes are designed and built, including the Square Kilometer Array (SKA), it is useful to review some of the recent advances made in experimental design and data analysis.
In particular, this paper summarizes recent progress made by three SKA pathfinder radio interferometers (\autoref{fig:telescopes}): the Murchison Widefield Array (MWA), the Low Frequency Array (LOFAR), and the Hydrogen Epoch of Reionization Array (HERA). We highlight their distinct approaches to data analysis informed by their instrumental design, as well as some of their synergistic properties.
We detail their design specifications in \autoref{tab:specs}, including the planned design of SKA-Low, highlighting its order-of-magnitude increase in collecting area relative to the current pathfinder experiments.
Much of their shared technical challenges revolve around the fundamental systematics facing 21\,cm interferometric experiments.\footnote{Note that the interferometric telescopes discussed in this paper are distinct from 21\,cm experiments targeting the ``global signal'' or the monopole signal \citep[e.g.][]{Bowman2018}, although they share many similar technical challenges.}
Largely, these challenges distill down to celestial, terrestrial, and instrumental systematics that impede the observers ability to separate foreground signals from 21\,cm signals, including:
\begin{itemize}
    \item {\it incomplete sky models:} Intrument calibration assumes that the sky model is known and perfectly represented. However, sky models are generally made from a catalogue of point sources, derived from previous observations with a given sensitivity and angular resolution, which is inevitably incomplete. Missing sources in the calibration model lead to spurious emission, often below the noise level \citep{Grobler2014}, that corrupts the foreground spectral smoothness, contaminating the EoR window \citep{barry_calibration_2016,Ewall-wice2017}. Moreover, radio sources exhibit a rich morphology, with jets and lobes that often extend over a range of angular scales: observations that fail to sample the full morphology inevitably lead to a calibration bias that, again, leaks foreground power into the EoR window \citep{Procopio2017}. 
    
    
    
    \item {\it primary beams:} Unlike steerable dish antennas, dipole arrays are mostly constituted of dipole clusters that are digitally pointed to a sky direction, resulting in primary beams that vary with time, which can be non-trivial to model. Additionally, primary beams have an intrinsic frequency dependence, particularly in the sidelobe region. Unmodeled variability in the primary beam sidelobes will inevitably corrupt off-axis foreground spectral smoothness and therefore leak into the EoR window \citep[e.g.,][]{Thyagarajan2016,Joseph2020,Nasirudin2020,trott_deep_2020}. Moreover, mutual coupling tends to complicate the spatial and spectral beam structure, exacerbating beam-to-beam differences \citep[e.g.,][]{Fagnoni2021}. Measuring, modeling, and mitigating primary beams effects is at the forefront of active research \citep{Orosz2020,Nunhokee2020,chokshi_dual_2021}, as they introduce biases that are difficult to correct for both standard and redundant calibration approaches.
    
    \item {\it ionosphere:} The partially ionized layer located between $\sim 50$ and 1000~km above the surface of the Earth interacts with the celestial signal at low frequencies. Absorption, scattering, and defocusing of the incoming wavefronts are due to temporal and spatial variations of the ionospheric electron density content, leading to a sky position and time-dependent point spread function. Ionospheric effects are routinely mitigated by building up a sky model and subsequently calibrating each sky source independently. This approach produces virtually artefact-free images \citep{Yatawatta2013,vanWeeren2016}, at the cost of extra computation. Theoretically, small offsets which lead to imperfect ionospheric calibration may lead to negligible contamination to short baselines \citep{Vedantham2016}. Fortunately, evidence suggests that most data are not severely affected by ionospheric effects at EoR frequencies \citep{jordan_characterization_2017}. However, the accidental inclusion of ionospherically-active data can obscure the 21\,cm signal \citep{trott_assessment_2018}. Ionospheric effects are dependent on $\lambda^2$, and thus CD analyses are more affected and must perform more mitigation \citep{yoshiura_study_2019}. 
    

    \item {\it baseline dependent effects:} Some effects cannot be factored and modeled by antenna-based terms and, therefore, cannot be treated on a per-antenna basis. Two examples of such errors are cross-coupling and radio frequency interference (RFI). Cross-coupling can group together a number of instrumental effects ranging from primary beams affected by mutual coupling, and actual cross-talk along the radio frequency system \citep{Kern2020b,Fagnoni2021}. Mitigation of some of these effects can be done in hardware \citep{Zheng2014}, while others can be carried out in an semi-empirical fashion \citep[e.g.,][]{Kern2020a}, as detailed physical modeling requires an extremely accurate knowledge of the whole signal chain and is different for each instrument.
    
    RFI is a well known problem for radio observations and advanced methods have already been developed to deal specifically for low frequency observations where the RFI environment is often more severe \citep[e.g.,][]{Offringa12,Offringa2015,Kerrigan2019}. Very faint RFI that cannot be detected and excised directly in the visibility data may be a significant contaminant to the EoR signal, and also lead to a calibration bias. Techniques to identify and remove data affected by faint RFI are actively being developed \citep{wilensky_quantifying_2020}.
   \end{itemize}

\begin{table}[h]
\begin{threeparttable}
\caption{SKA pathfinder experiment specifications.}
\vspace{0pt}
\begin{tabular}{| p{5.5cm}  p{2.2cm}  p{2.2cm}  p{2.2cm} p{2.2cm} |}
\hline
\rowcolor{gray!10}
Parameter & MWA$^a$ & LOFAR \citep{vanHaarlem13} & HERA \citep{DeBoer2017} & SKA-Low$^b$ \\
\hline
\hline
Latitude, longitude [$^\circ$] & -26.7, 116.6 & 52.9, 6.9 & -30.7, 21.4 & -26.8, 116.8   \\
\hline
\rowcolor{gray!10}
Element size [m] & 5 & 30.75 & 14 & 38 \\
\hline
Number of elements & 128 & 76 & 350 & 512 \\
\hline
\rowcolor{gray!10}
Collecting area [m$^2$] &  $\mathcal{O}(10^3)$ & $\mathcal{O}(10^4)$ & $\mathcal{O}(10^4)$ & $\mathcal{O}(10^5)$ \\
\hline
Min. baseline length [m] & 7.7 (7.1) & 68 & 14.6 & 40\\
\hline
Max. baseline length [m] & 2873 (687) & $2 \times 10^6$ & 876 & 65000\\
\hline
\rowcolor{gray!10}
Angular resolution$^\ast$ [arcmin] & 2 (6) & $3 \times 10^{-3}$ & 11 & 0.15\\
\hline
Field of view$^\ast$ (FWHM)[$^\circ$] & 25 & 3.8 & 9 & 3 \\
\hline
\rowcolor{gray!10}
Frequency coverage [MHz] & 80--300 & 30--190 & 50 -- 230 & 50 -- 350\\
\hline
\end{tabular}
\vspace{0em}
\begin{tablenotes}
  \footnotesize
  \item[$\ast$] Angular resolution and field of view quoted at 150 MHz.
  \item[a] Phase I and (Phase II) MWA EoR layout.
  \item[b] Planned design as of \cite{Braun2019}.

\end{tablenotes}
\end{threeparttable}
\label{tab:specs}
\end{table}

\begin{figure*}
\centering
\includegraphics[width=\linewidth]{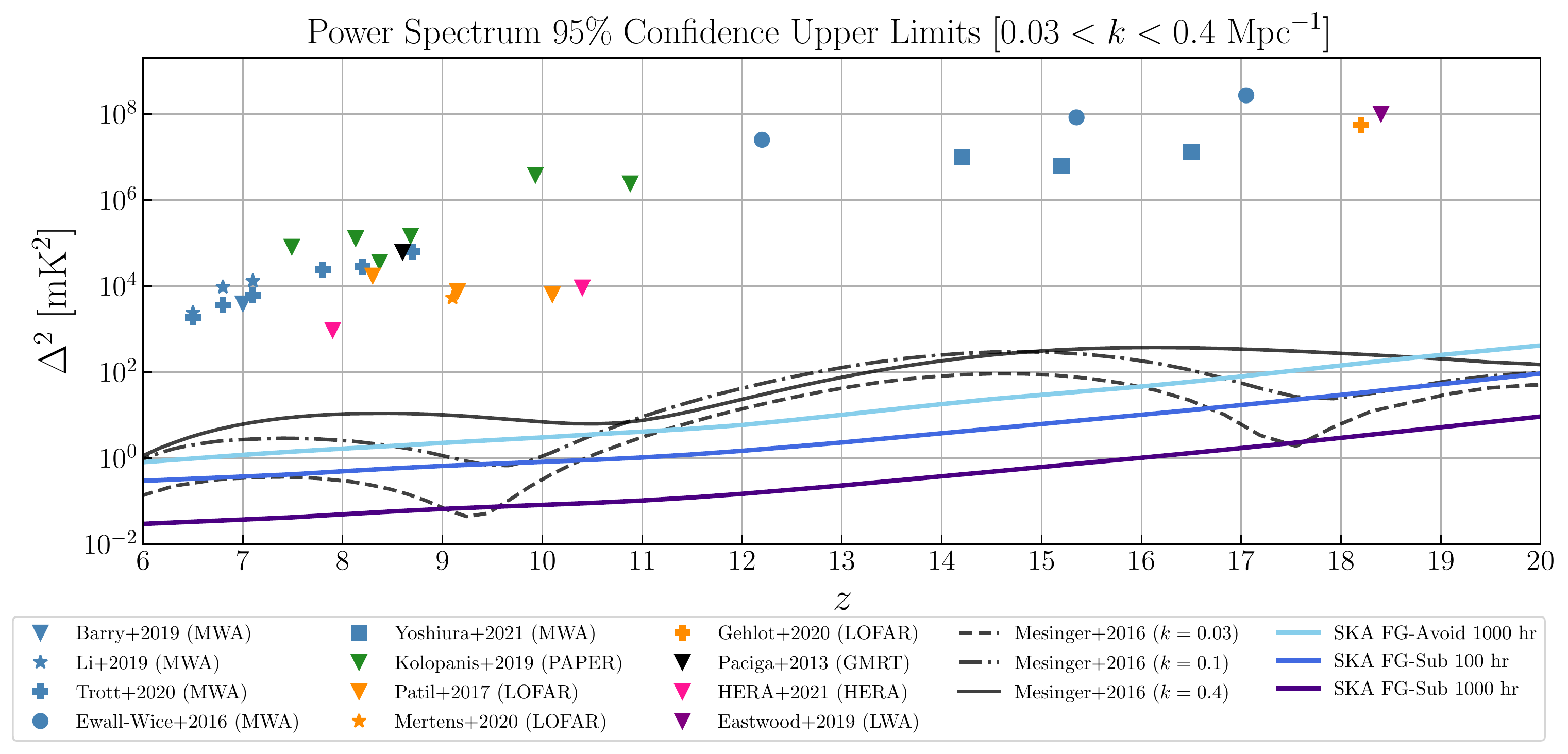}
\caption{Upper limits on the 21\,cm power spectrum at 95\% confidence ($2\sigma$) from various experiments from $6<z<20$ spanning a range of wavevectors, $k$. The redshift range is chosen to focus on recent limits from SKA pathfinders. The theoretical power spectrum from the faint galaxies EoR simulation of \cite{Mesinger16} is plotted as solid and dashed black lines. While 21\,cm interferometric experiments have steadily pushed down in sensitivity over the past five years, fiducial models remain a couple orders of magnitude deeper. Projected $2\sigma$ sensitivity curves for the SKA assuming foreground avoidance at $k=0.4 {\rm Mpc}^{-1}$ (FG-Avoid) and foreground subtraction at $k=0.1\ {\rm Mpc}^{-1}$ (FG-Sub) are also plotted for a 100 hour and 1000 hour integration\protect\footnotemark.}
\label{fig:limits}
\end{figure*}

\footnotetext{The SKA-low sensitivity is computed using the latest available stations layout as of July 2021 and latest single station sensitivity estimates~\citep{deLera20}.}

\begin{figure*}
\centering
\includegraphics[width=\linewidth]{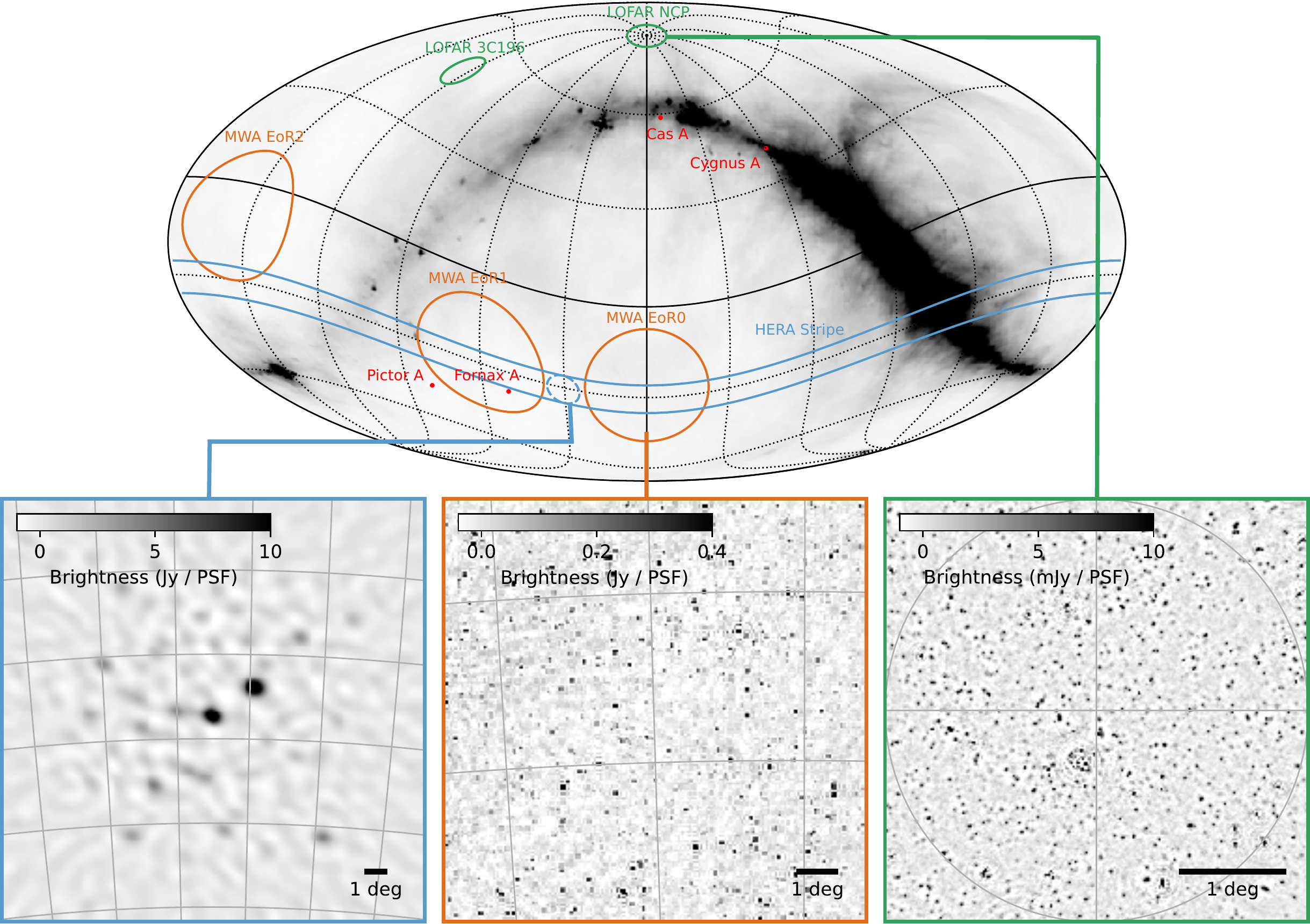}
\caption{Images of the fields observed by the three SKA pathfinder experiments. At the top, we show a map of the low-frequency galactic foregrounds \citep{Oliveira2008, Zheng2017}, highlighting the main observing fields of HERA (Phase I; blue), MWA (Phase I; orange), and LOFAR (green), as well as the locations of a few prominent radio sources (e.g. Fornax A). At the bottom, we show representative images taken by each of the experiments illustrating the distinctive fields of view and spatial resolutions \citep[reproduced with permission from ][]{Kern2020b, barry_improving_2019, Mertens20}.}
\label{fig:fields}
\end{figure*}

Recent developments in addressing some of the aforementioned systematics, some at the hardware level others at the analysis level, have led to improved constraints on the upper limit of the 21\,cm power spectrum at the EoR and CD, summarized in \autoref{fig:limits}.
We also plot fiducial power spectrum models, showing a few orders of magnitude in sensitivity still needed in order to make a fiducial detection.
We also plot forecasted noise curves from the SKA under different time integration and foreground mitigation assumptions, showing it will be able to detect these models across a wide range of redshifts at high significance, assuming systematics can be mitigated. \autoref{fig:fields} shows a map of the low-frequency sky, highlighting parts where the SKA pathfinders have set their most sensitive upper limits on the power spectrum, with representative images of those foregrounds with each of the experiments.

In the following sections, we review improved limits from the MWA, LOFAR, and HERA, and discuss some of the developments that enabled these improvements. In addition, we discuss some of the challenges in theoretical simulation of the 21\,cm signal, and the interpretation of these upper limits for placing constraints on astrophysical models of the EoR and CD.

\section{The Murchison Widefield Array}
\label{sec:mwa}

The Murchison Widefield Array\footnote{\url{http://mwatelescope.org}} \citep[MWA;][]{tingay_murchison_2013, wayth_phase_2018} is a first-generation low-frequency radio interferometer located in the Murchison Shire of Western Australia. The MWA is an observatory; it has a wide-range of use-cases ranging from extragalactic studies of clusters and AGN, to nearby studies of our own Galaxy and Sun. One of its main scientific objectives, however, is to measure the elusive 21\,cm signal from our early Universe, and has been observing three low-foreground fields on the sky (EoR0, EoR1, and EoR2; see~\autoref{fig:fields}) in search of the EoR and CD since 2013. 

In order to reduce the number of observations required to reach EoR sensitivities, ultra-wide field-of-views are built into the design of the MWA. This is achieved through the use of beamformed elements, consisting of a 4x4 grid of dual-polarisation dipoles over a ground screen. The frequencies of interest for EoR observations with the MWA range from 139 to 197\,MHz \citep[e.g.][]{trott_deep_2020} which probes $z=9.3-6.2$, but earlier epochs can be observed up to $z=16.5$ \citep{yoshiura_study_2019}. 

\subsection{Technical advancements}
\label{sec:mwa:techincal_advacements}

The MWA has been undergoing various upgrades and transformations throughout its history. Phase I of the MWA \citep{tingay_murchison_2013}, from 2013 to 2017, consisted of 128 beamformed elements arrayed in a semi-random layout with a dense core to allow for high-fidelity imaging and large-scale sensitivity. Phase II of the MWA \citep{wayth_phase_2018}, from 2017 to 2022, has two configurations depending on the observing program: one imaging layout with longer baselines, and one redundant layout with a subset of shorter, random baselines (see~\autoref{fig:telescopes}, orange insert), each with 128 beamformed elements. The addition of redundancy has allowed for extensive testing of redundant-based analysis techniques in direct comparison with standard imaging techniques \citep{kerrigan_improved_2018,li_comparing_2018,zhang_impact_2020}, especially in preparation for the next scheduled upgrade.

Phase III of the MWA will be an extensive upgrade of both digital and analogue systems, including new beamformers, cables, receivers, and correlator. In the context of EoR, the upgrades to the digital and analogue systems are being driven by the difficulties in calibration. Removing the instrumental signal from the data itself needs to be done at extremely high precision and accuracy, at least one part in 10$^5$ for both sky-based \citep{barry_calibration_2016} and redundant-based \citep{byrne_fundamental_2019} calibration schemes. Given this requirement, studies on the stability of the beam have begun in earnest to characterise the effects of instrument degradation \citep{chokshi_dual_2021} or deviation \citep{joseph_bias_2018} on the EoR signal.

These upgrades affect almost every stage of the measurement process, from the element to the correlator. Firstly, the coaxial cables from each antenna to the beamformer box will need to be replaced. Over the course of a decade, the coaxial shielding has degraded in the harsh environment of the Outback, resulting in a sensitivity loss. Next, the receivers will be completely replaced, including the digital system and the analogue signal conditioning (ASC) system. An oversampled polyphase filterbank will be used to eliminate loss and spectral artefacts caused by coarse channelisation, and when included with increased bit depth for a larger linear-response regime, the EoR spectral requirement of one part in 10$^5$ can be reached. Improvements to the ASC will also allow the MWA to measure down to 50\,MHz, making it a more effective CD experiment. Finally, the MWAX correlator will be able to correlate at least 256 tiles, thus combining the two observing styles using the oversampled coarse channel inputs. 

\subsection{Analysis advancements}
\label{sec:mwa:analysis_advancements}

Integrating petabytes of data to reach EoR-levels of sensitivity is compounded in difficulty by the necessary precision and accuracy of the analysis. Various studies have been done within the context of the MWA to understand and reduce systematics, or analysis-based errors, in preparation for the future SKA. These can be grouped into two broad catagories: foreground mitigation and precision analysis. 

RFI, particularly FM/AM radio and digital TV, is an unwanted foreground and thus must be removed from the data. Even though the MWA is located on a designated radio-quiet site with low RFI background \citep{offringa_low-frequency_2015}, unchecked faint RFI can still contaminate the EoR signal \citep{wilensky_quantifying_2020}. This is evident in a recent upper limit on the EoR power spectrum from the MWA, where mitigating ultra-faint digital TV \citep{wilensky_absolving_2019} resulted in a factor of 3 improvement \citep{barry_improving_2019}. 

The refractive effects of the ionosphere can translate into contamination on the EoR signal, and thus must be mitigated as well. By using the severity of the positional offsets of point-source foregrounds as a metric for activity \citep{jordan_characterization_2017}, we can now determine the error budget of unmitigated ionospheric effects on EoR analyses \citep{trott_assessment_2018,chege_simulations_2021}. These techniques have been used in recent EoR power spectrum upper limits from the MWA to reduce ionospheric systematics \citep{trott_deep_2020}, and become vitally important for higher redshift measurements \citep{yoshiura_study_2019} due to the wavelength dependence of the ionospheric offsets.

Whilst foreground avoidance techniques do not inherently require subtraction due to the modes of interest, they stand to benefit from an improvement in dynamic range. Thus, even though the MWA probes EoR-window modes, much effort has gone into the characterisation of foregrounds, recently via diffuse mapping \citep{byrne_map_2021} and with wavelet decomposition of complicated source morphologies \citep{line_modelling_2020}. 

One of the most dramatic improvements in MWA EoR analysis has been through extreme precision and accuracy, specifically through enforcing spectral smoothness throughout all software modules. Calibration in particular has been a main focus for the MWA, both in discovering spectral requirements \citep{barry_calibration_2016,byrne_fundamental_2019} and investigating/inventing various calibration styles \citep{li_comparing_2018,barry_fhd/eppsilon_2019,zhang_impact_2020}. Recently, enforcing spectral smoothness via the instrumental Fourier kernel function through a Tapered Gridded Estimator \citep{choudhuri_tapering_2016} and with extreme overresolution \citep{Offringa19a} resulted in significant improvements in MWA upper limits \citep{barry_improving_2019,li_first_2019, trott_deep_2020}. Given the importance of spectral smoothness to the power spectrum measurement, current efforts focus on effects of degradation on the instrumental beam \citep{chokshi_dual_2021} and how that propagates to the power spectrum.

In order to explore observing methodologies and their corresponding analysis techniques, the MWA can observe a singular field (like LOFAR) or allow the sky to pass overhead (like HERA). Most published limits with the MWA data come from single-field observations, but investigations into how these results compare to a drift-scan observing programs have shown promising results. First attempted in 2016 \citep{paul_delay_2016}, these analyses are now comparable with previously published limits \citep{patwa_extracting_2021}, showing significant improvement in precision techniques for various analysis methodologies with the MWA. 

\subsection{Limits and future developments}
\label{sec:mwa:future_developments}

The MWA has a crucial role in SKA development as a pathfinder for the low-frequency program. Indeed, much of the future development of the MWA, including technical and software advancements, is in preparation for the SKA. 

Besides the Phase III upgrade already underway, an SKA-MWA signal bridge for cross-correlation is being developed. This will be essential to the EoR experiment due to the harsh spectral requirements on response and calibration. By being able to correlate MWA elements with the SKA prototypes, we can have a stronger view on the future performance of these prototypes, and thus be able to change undesired responses prior to the full build. This does come with challenges; the sample rate is unmatched between the MWA and the prototypes, thus a resampling and rate conversion signal bridge will need to be developed alongside Phase III specifications.

Teams around the world are also focusing on analysis software as we approach the SKA-era. Pipelines with various degrees of complexity are being developed, allowing for simultaneous comparisons and cross-pipeline outputs \citep{jacobs_murchison_2016}. More complex pipelines will handle the scaling and data load that the SKA will bring \citep[e.g.,][]{mitchell_real-time_2008}, whilst simpler pipelines are the sandbox of scientific testing \citep[e.g.,][]{barry_fhd/eppsilon_2019}. As these pipelines are improved, the MWA collaboration publishes upper limits using our current data stores to document our progress. 

In 2019, a re-analysis of 21\,hrs of Phase I MWA data resulted in a power decrease of almost a factor of 10 \citep{barry_improving_2019}, owing to careful removal of ultra-faint RFI and improved precision analysis techniques. This was surpassed by an analysis of Phase II MWA data using 40\,hrs of data, which combined both redundant and sky-based analysis techniques to further reduce systematics \citep{line_modelling_2020}. In 2020, the best-selected 110\,hrs of MWA data resulted in an upper limit of  $\Delta^2_{21} < (43)^2\,\mathrm{mK^2}$ at $k = 0.14\,h\mathrm{Mpc^{-1}}$ and $z=6.5$ \citep{trott_deep_2020}, which is to-date the best limit from the MWA. However, a variety of redshifts have been probed by various works with the MWA, including CD redshifts \citep{ewall-wice_first_2016,yoshiura_study_2019}, as shown by~\autoref{fig:limits}.

\section{The Low Frequency Array}
\label{sec:lofar}

\subsection{Project Overview}

LOFAR\footnote{Low-Frequency Array, \url{http://www.lofar.org}}~\citep{vanHaarlem13} is a radio telescope with its core in the Netherlands and several ``international stations'' located in multiple European countries. It consists of two distinct aperture arrays: the Low-Band Antenna (LBA) system (30-90 MHz) and the High-Band Antenna (HBA) system (110-190 MHz). The observation of the 21\,cm signal from the EoR has been one of the drivers behind the design of the telescope: the combination of a dense core array, providing great sensitivity at large spatial scale, and long to very long baselines, allowing for very high-resolution and deep imaging, make it an ideal instrument for precision foreground characterization and subtraction for detection of the cosmological signal.

The LOFAR-EoR Key Science Project (KSP) mainly targets two deeps fields: the North Celestial Pole (NCP) and the field surrounding the bright compact radio source 3C\,196 (see~\autoref{fig:fields}). Currently $\approx 2480$ hours of data have been observed with the HBA system. So far, the deepest upper limits of the instrument have been observed with the NCP field. The team published in 2017~\citep{Patil17} its first upper limit at $z=8.3, 9.1$ and 10.1 from one night of observation, and in 2020~\citep{Mertens20}, an improved upper limit at $z=9.1$ combining 141 h of data. Step by step, progress has been made, but the experiment remains very challenging. Many studies were needed, and still are, to control the many complex aspects in the signal processing chain in order to be able to reach the expected 21\,cm signal strengths which lie two to three orders of magnitude below current upper-limits. Contaminants have many sources: gain-calibration errors due to an incomplete or incorrect sky model~\citep{Patil16}, errors in band-pass calibration, beam modeling, residual RFI~\citep{Offringa19a}, as well as chromatic errors introduced due to leakage from the polarized sky into Stokes I~\citep{Jelic10,Asad15} or due to the ionosphere~\citep{Mevius16}. These challenges can be met for a large part by a constant improvement of the processing pipeline.

The LOFAR-EoR data processing pipeline consists, in essence, of (1) Pre-processing and RFI excision, (2) direction-independent calibration (DI-calibration), (3) direction-dependent calibration (DD-calibration) including subtraction of the sky-model, (4) imaging, (5) residual foreground modelling and removal, and (6) power spectra estimation. All data processing is performed on a dedicated compute-cluster called Dawn~\citep{Pandey2020}, which consists of 48 $\times$ 32 hyperthreaded compute cores and 124 Nvidia K40 GPUs.

\subsection{Analysis advancements}

The latest LOFAR published upper limit~\citep{Mertens20} was achieved by making substantial progress in mainly two areas of the processing pipeline: DD-calibration and residual foreground suppression. 

LOFAR has a wide field-of-view (about 10$^\circ$ between primary beam nulls at 140 MHz) and is susceptible to many direction-dependent effects (mainly primary beam and ionospheric effects). Appropriate gain solutions in the source directions are needed to subtract them. Solving in the direction of each and every sources of the sky model would be impractical. Sagecal-CO~\citep{Yatawatta16} is used to solve gain solutions simultaneously in 122 clusters of the NCP sky model (which consists of 28755 components), but additional constraints need to be set to make the problem tractable. To further reduce the number of degrees of freedom, spectral smoothness of the instrumental gains is imposed through strong regularisation, which has the combined advantage of reducing calibration noise and making the calibration more resistant to signal loss~\citep[]{Sardarabadi19}(Mevius~et~al.~in prep.).

The second area where substantial progress has been made is in the residual foreground suppression strategy. The LOFAR-EoR project relies heavily on the foreground removal paradigm to access the largest available scales of the power-spectra, for which LOFAR is most sensitive. Since the inception of the project, the team has developed and tested several methods to model and subtract foreground emissions: polynomial fitting~\citep{Jelic08}, Wp smoothing~\citep{Harker09}, FastICA and GMCA~\citep{Chapman13}. Most of these methods exploit the distinctive spectral correlation signature of the various constituents of the observed signal. While these methods have performed very well on simulated data, their application to real observations has proved more difficult~\citep{Hothi21}. A new foreground removal algorithm based on Gaussian Process Regression (GPR)~\citep{Mertens18} is now used, which tries to overcome some of the shortfalls found in earlier techniques. In this method, a statistical model of all the components contributing to the observed signal is built by using parametric covariance functions. This formulation ensures a relatively unbiased separation of their contribution and an accurate estimation of the uncertainty.

Combining in total 140 hours of observation on the NCP at $z=9$, an upper limit on the 21-cm power spectrum of $\Delta^2_{21} < (72.86)^2\,\mathrm{mK^2}$ at $k = 0.075 \mathrm{h\,cMpc^{-1}}$ was obtained. This has resulted in a reduction by a factor 10 from our previous upper limit at this redshift~\citep{Patil17}. This is the deepest upper limit at $z=9$, but is not yet optimal as it is affected by a large excess of power, particularly at large scales where the observed residual power is an order of magnitude brighter than the thermal noise power. Nevertheless, this is still an astrophysically interesting upper limit, and has been used in several publications to set constraints on the physics of the EoR and on the level of excess radio background (see~\autoref{sec:interpretation}).

\subsection{Current and Future Developments}

Reaching deeper upper limit requires tackling the issue of excess power. Being partially correlated between observing nights, this excess prevents the residual power from being reduced when they are combined. Its origin is investigated in detail in Gan~et~al. (in prep). In particular, taking the diffractive scale as a metric for ionospheric behavior~\citep{Mevius16}, no clear correlation was found  between the excess power and ionospheric activity. This challenges the idea that the ionosphere is a severe and limiting factor, although it might still be an issue at lower noise levels. On the other hand, by dividing the observations into 3-hour Local Sidereal Time (LST) bins, it was found that the level of excess was very different from one LST bin to another, and that it was correlated with the flux of bright and distant sources. This suggests that the excess is sky-related, which could be directly (through beam side-lobes) or indirectly (e.g. through gain errors). As a result, several improvements at different stages of the processing pipeline have been planned. In particular, efforts are focused on the calibration steps, but also on the elimination of faint and broadband radio interference, as well as on the suppression of residual foregrounds.

In detail, to reduce calibration noise to a minimum, and prevent gain solutions from absorbing flux from sources not part of our sky model,  DI-calibration is split in two, with a first step in which spectral-smoothness of the calibration gains is fully enforced and a second step which captures the fast frequency varying but time stable band-pass response of the instrument. While the improvement in the resulting power-spectra is only marginal, this step is now much more resilient to signal loss~\citep{Sardarabadi19}. 

For the DD-calibration step, spectral-smoothness of the calibration gains is also being fully enforced now, and the sky model is being gradually improved with a focus on the brightest sources in the sky (Cassiopea A and Cygnus A), and on the sources located in the second side-lobe of the LOFAR primary beam. The resulting reduction in residual power is substantial, reaching a factor ~10 in some parts of the transverse vs line-of-sight power-spectra.

RFI is also a major concern in 21\,cm experiments, and in particular faint and broadband RFI which are tricky to detect and filter. The technique of near-field imaging~\citep{Paciga11} is used to locate local sources of RFI and affected baselines are then flagged accordingly. An additional AOflagger~\citep{Offringa12} step after DD-calibration is also added. Significant improvements is noted in the cylindrically-averaged power-spectra along the horizon line and at low $k_\perp$ (small baselines).

Finally the residual foreground removal algorithm is being revised to more optimally separate the 21\,cm signal from the foregrounds and to make it more robust against signal loss. The parametrized covariance model for both the 21\,cm signal and foregrounds is enhanced by incorporating information from simulations via machine learning. The covariance description is also extended to the spatial domain. This is still work in progress but tests on simulation have already shown very promising results.

Despite the challenge of the experiment, the LOFAR-EoR project has made progress towards a detection of the 21\,cm from the EoR in recent years and is currently preparing a new multi-redshift upper limit combining the many improvements introduced in the processing pipeline. The LOFAR 2.0 project is also underway. The first stage of upgrades will significantly increase the sensitivity of the LBA system, opening a new window of opportunity to explore the CD with LOFAR, in addition to the ACE project~\citep{Gehlot20} already in operation and which uses the AARTFAAC system of LOFAR.


\section{The Hydrogen Epoch of Reionization Array}
\label{sec:hera}

\subsection{System Overview}

The Hydrogen Epoch of Reionization Array\footnote{\url{http://reionization.org/}} (HERA) is a purpose-built radio interferometer for 21\,cm intensity mapping of the EoR and CD.
It is based in the South African Karoo Astronomy Reserve \citep{DeBoer2017}, and when completed will consist of 350 dish antennas (320 in a dense core with 30 outrigger antennas) packed in a highly redundant and compact hexagonal configuration observing from 50 -- 225 MHz.
\autoref{fig:telescopes} shows a satellite view of the array as of early 2021 (blue-border inset), showing the completed dish deployment of the HERA core.

Relative to configurations like the MWA and LOFAR, which are at least partially optimized for point source imaging,  HERA's redundant and compact configuration make it less suitable for high angular resolution imaging, often needed for precise foreground characterization.
Instead, HERA's configuration maximizes its sensitivity to the 21\,cm power spectrum and allows it to exploit its high degree of baseline redundancy for alternative calibration techniques like redundant calibration \citep{Dillon2016, DeBoer2017, Dillon2020}.
This difference in array configuration highlights the stark contrast in how various experiments aim to deal with the foreground contamination problem: either through modeling and subtraction, or through avoidance.
Indeed, central to HERA's design is the idea of foreground containment and avoidance.
A drift-scan array, HERA does not have any moving parts involved with observations, meaning its front-end response is generally quite stable over time.
Furthermore its compact array configuration keeps intrinsic foreground contamination to largely smooth spectral modes, thus preserving more modes for measuring the cosmological signal \citep{Parsons2012}.
Challenges to the foreground avoidance strategy HERA is pursuing include high dynamic range calibration of the antenna response, and the mitigation of low-level instrumental systematics like mutual coupling and cross-talk.

HERA's construction has progressed in two phases.
Phase I utilized the existing on-site infrastructure from HERA's predecessor, the Precision Array for Probing the Epoch of Reionization (PAPER) experiment \citep{Parsons2010}, such as its dipole feeds and signal chains, and combined them with new HERA dishes, which significantly increased HERA's total collecting area compared to PAPER.
During Phase I construction of the core of the array (from 2017 to 2018), observations were conducted at night for commissioning and data analysis purposes.
A subset of this data forms the basis of HERA's first limit on the 21\,cm power spectrum \citep{hera2021}.
More recently, HERA construction has progressed to Phase II, which notably involves the installation of new wideband feeds \citep{Fagnoni2020_wideband}, in addition to the replacement of the front-end signal chain and an upgrade to the correlator \citep{DeBoer2017}.
To-date, all of the 330 antennas in the core have been constructed (\autoref{fig:telescopes}), with roughly half of them having the new feed and signal chains installed.
Current on-site work is focused on installing the remaining feeds and commissioning the system, with full science observations projected for 2022.

\subsection{Advances in Instrument Modeling}
\label{sec:hera_modeling}

Recent advances in modeling the response of the HERA antenna, feed, and signal chain have come from detailed electromagnetic simulations \citep{Fagnoni2021, Fagnoni2020_wideband}, which include models for the direction-dependent and direction-independent response of both isolated HERA antennas and antennas embedded in a compact array.
The predicted beam models are at least partly confirmed by in-situ beam modeling derived from imaging drift-scan point source tracks \citep{Nunhokee2020}, which shows that the beam is well matched by observations out to the first sidelobe.
Current and ongoing work is focused on improving existing models of the far sidelobes and understanding their impact on the 21\,cm power spectrum \citep[e.g.][]{Choudhuri2021}.

The embedded-element simulations of \citep{Fagnoni2021} also predict the presence of antenna-to-antenna coupling in the HERA system that may be more complex than originally considered.
Indeed, analyses of Phase I data reveal the presence of baseline-based (rather than antenna-based) instrumental coupling systematics \citep{Kern2020a, Thyagarajan2020}.
While such systematics make detection of the 21\,cm signal more difficult, they can be at least partially mitigated.
For example, simulated and empirical modeling of the HERA Phase I system by \citep{Kern2019, Kern2020a} shows that antenna-based models for cable reflections and baseline-based, time-domain Fourier filters for antenna coupling can mitigate the observed instrumental systematics by at least two orders of magnitude in the power spectrum.

A deeper understanding of HERA's instrumental response has also come from the development of HERA's calibration pipeline.
The 21\,cm foreground challenge necessitates that spurious frequency and time structure in the antenna gain solutions be minimized in order to ensure a clean separation of the cosmological signal.
For HERA, the dominant effects are thought to be imperfect models of the sky at the fields used for calibration, as well as deviation from ideal redundancy between nominally redundant baselines.
Studies of HERA Phase I data do show evidence for such spurious structures, which have yielded insights into the performance of the system, such as the impact of poorly modeled diffuse foreground emission \citep{Kern2020b}, and the nominal redundancy of HERA baselines \citep{Dillon2020}. Such effects will hinder recovery of the 21\,cm power spectrum; however, studies indicate they can be at least partially mitigated by smoothing the calibration solutions across time and frequency with tailored Fourier domain filters \citep{Kern2020b, Dillon2020}.

\subsection{Improved Power Spectrum Limits at $z=7.9$}
\label{sec:hera_limits}

Recently, a full analysis of an 18-day dataset from HERA Phase I was completed ($\sim$30 hours of integration), yielding new upper limits on the 21\,cm EoR power spectrum at $z=10.4$ and 7.9 \citep{hera2021}.
These limits are $\Delta^2_{21} \le (30.76)^2\ {\rm mK}^2$ at $k=0.192\ h\ {\rm Mpc}^{-1}$ and $z=7.9$, and $\Delta^2_{21} \le (95.74)^2\ {\rm mK}^2$ at $k=0.256\ h\ {\rm Mpc}^{-1}$ and $z=10.4$ at 95\% confidence \citep{hera2021}. These limits are compared against existing limits in \autoref{fig:limits}, showing how the $z=7.9$ limit by HERA improves upon existing limits by over an order of magnitude.
The deepest limits achieved a dynamic range of $10^9$ with respect to the peak measured foreground power, which was enabled by HERA's emphasis on a smooth instrumental response and a careful accounting of spectral leakage in their analysis pipeline.
Furthermore, contrary to other recent limits from competing experiments, HERA's deepest upper limits exhibit rough consistency with thermal noise fluctuations for $k\ge0.2\ h\ {\rm Mpc}^{-1}$, suggesting that the limits could be further improved with more data.
This conclusion was derived from a number of statistical null tests that carefully compared the data against HERA's noise models, testing analysis choices such as the frequency window weighting function and the selection of observing nights integrated in the final dataset \citep{hera2021}.

The results presented by \citep{hera2021} also benefited from a detailed quantification of the uncertainty on the measured power spectrum.
\cite{Tan2021} provides an overview of the different error-bar methodologies explored by HERA, examining the strengths and weaknesses of various analytic and empirical methods.
The chosen methodology for reporting upper limits in \citep{hera2021} is able to both robustly estimate the thermal noise floor of the data, and can also account for boosted noise fluctuations sourced by residual systematic cross terms.

To further bolster confidence in the power spectrum limits set by \citep{hera2021}, an independent validation effort was undertaken to ensure that the HERA Phase I analysis pipeline could recover a known input signal from a realistically corrupted mock HERA data simulation \citep{Aguirre2021}.
This effort sought to validate many facets of the analysis pipeline, including the accurate simulation of wide-field foreground and EoR signals, the unbiased recovery of direction-independent gains, the mitigation of baseline-based systematics, and the unbiased estimation of the 21\,cm power spectrum.
Notably, it tested these components both as individual blocks isolated from other parts of the pipeline, and as an end-to-end chain where the pipeline is run on the simulated ``raw'' data all the way to power spectrum estimation.
One of the more tangible outcomes of the validation analysis was the discovery of a $\sim7\%$ overall bias in the flux scale of the HERA calibration pipeline, which was then corrected.
While the validation effort of \citep{Aguirre2021} was ambitious in terms of the number of pipeline components tested, it also laid the framework for increasingly more comprehensive validation efforts that will be applied to future HERA results.

Some of the current challenges for HERA data analysis include modeling and mitigating the impact of poorly understood diffuse foregrounds \citep{Thyagarajan2016, Carilli2020}, mitigating residual antenna-based and baseline-based instrumental systematics, and identifying weak levels of radio frequency interference in the data \citep{Kerrigan2019, wilensky_absolving_2019}. 
Examples of recent and ongoing work focused on meeting these challenges include the design of more sophisticated models of instrumental coupling systematics, and the constructing visibility filters that remove such systematics to higher dynamic range \citep[e.g.][]{Ewall-Wice2021}.
Current analysis efforts are focused on applying the HERA calibration and power spectrum estimation pipeline on additional Phase I data to achieve possibly deeper limits across $6 < z < 11$, as well as applying it to more recent HERA Phase II data to improve existing power spectrum limits at $z > 12$.

\section{Theoretical Modeling}
\label{sec:theory}

The primary goal for the theoretical modelling of the 21\,cm signal is to be able to infer the underlying physics driving the EoR and CD as revealed by the observations. For example, we want to be able to extract information about the nature of these first galaxies (e.g. escape fraction of ionising radiation). To infer this information we require three major steps: \textit{(i)} the ability to characterise the observational data (e.g. with a summary statistic), \textit{(ii)} a theoretical model for the 21\,cm signal and \textit{(iii)} a probabilistic framework from which to infer astrophysical properties using \textit{(i)} and \textit{(ii)}.

\subsection{Characterising the 21\,cm signal}

The 21\,cm signal observed by radio interferometers varies both spatially (transverse to the observers line-of-sight) and in frequency (along the observers line-of-sight). Thus, measuring the 21\,cm signal yields a 3D movie that reveals the timeline of the Universe and contains a wealth of both astrophysical and cosmological information.  

The sheer volume of data expected is so vast that we cannot conceivably utilise each independent piece of information. Instead, we must reduce the data into a more manageable data-set, typically achieved by statistically averaging the signal using some metric or summary statistic. For studying the EoR, the most commonly adopted and heavily studied statistic in the literature has been the 21\,cm power spectrum \cite{Pober:2014,Greig:2015,Hassan:2017,Cohen:2018}. This is primarily because interferometers natively observe in Fourier space and that the first pathfinders likely can only achieve a low signal-to-noise detection. However, the 21\,cm signal is highly non-Gaussian owing to the complex 3D ionisation morphology arising from the overlap of distinct ionised regions. Thus, the 21\,cm power spectrum is not an optimal statistic to describe the EoR as it crucially misses the non-Gaussian information. 

Recently, with the expected order of magnitude increase in sensitivity achievable with the forthcoming SKA, considerable effort has been spent exploring more optimal 21\,cm statistics with a specific focus on accessing the non-Gaussian information. The most prominent of these has been the 21\,cm bispectrum \cite[e.g.][]{Yoshiura:2015,Shimabukuro:2017,Majumdar:2018,Hutter:2019,Watkinson:2019}: the natural extension of the 21\,cm power spectrum which can notably improve our ability to infer astrophysical information. Alternatively, non-Gaussian information can further be extracted using the Morlet transform \cite{Trott:2016} along with the higher order moments of the 21\,cm brightness temperature PDFs \cite[e.g.][]{Watkinson:2015,Kubota:2016}. Further, with the expected imaging capabilities of the SKA, it has also been shown that astrophysical information can equally be extracted from 2D tomographic images of the 21\,cm signal. Information can be gleaned about individual ionised bubbles or regions of interest using matched filters \cite{Datta:2012,Majumdar:2012} or convolutional neural networks \cite{Hassan:2019a}, the distribution of ionised regions using sophisticated image analysis techniques \cite{Kakiichi:2017,Giri:2018} or machine learning \cite{Bianco:2021}, the average properties of galaxies from stacking low-resolution images of the 21-cm signal centred on ionised regions \cite{Davies:2021} and the topological and morphological classifications of the 21-cm signal \cite[e.g.][]{Yoshiura:2017,Kapahtia:2018,Elbers:2019,Giri:2021}. Clearly, each approach has its own strengths and weaknesses and its performance is strongly tied to the specific astrophysical information or feature that is being investigated. It is worth noting that these are only a select few examples of the broach range of approaches being explored in the literature.

\subsection{Modelling the 21\,cm signal}

As the EoR and CD are a complex 3D radiative transfer problem, modelling the 21\,cm signal requires numerical simulations. However, current simulations do not have the dynamic range to self-consistently model the EoR (i.e. simultaneously resolve individual stars while tracking large-scale ionisation fronts). Instead, in our theoretical toolkit we have a suite of simulation techniques tailored to exploring specific physical questions trading physical accuracy for computational efficiency. These include coupled hydrodynamical and radiative transfer simulations ($\sim1-10$~Mpc's in size), which are extremely computationally intensive that primarily focus on self-consistently exploring first galaxy formation and resultant internal feedback mechanisms \cite[e.g.][]{Rosdahl:2018,Wu:2019,Ocvirk:2020}, and more recently \cite{Kannan2021, Garaldi2021, Smith2021}. Hydrodynamical or N-body simulations with radiative transfer applied in post-processing (typically $\sim10-100$~Mpc's in size) that are capable of capturing the moderate to large scale distribution of galaxies responsible for reionisation \cite[e.g.][]{Dixon:2016,Semelin:2017,Eide:2020}. Galaxy semi-analytic models coupled with semi-numerical simulations ($\sim100$~Mpc's in size) which focus on galaxy formation and evolution physics and how these processes imprint their signature on the large-scale 21\,cm signal \cite{Mutch:2016,Hutter:2021}. Finally, approximate but computationally efficient semi-numerical simulations \cite{Santos:2010,Mesinger:2011,Fialkov:2014,Hutter:2018} are used for rapid astrophysical parameter space exploration or generating extremely large cosmological volumes (e.g.  $\gtrsim1$Gpc).

\subsection{Inference of astrophysical information} \label{sec:interpretation}

The inference of astrophysical information on the EoR arises from our ability to compare our models of the 21\,cm signal to the observational data (through some metric or summary statistic as described above). Here, we want to quantify the likelihood that our theoretical models (given a set of input astrophysical parameters describing the physics) match the observation of the 21\,cm signal. Ideally, one would perform this in a fully Bayesian framework, however, this is only currently feasible with semi-numerical simulations (i.e.~to be able to rapidly explore the vast astrophysical parameter space). 

In recent years several tools have been developed specifically to perform Bayesian inference of the astrophysical parameters describing the EoR. These include \textit{(i)} flexible, direct approaches that simulate the full 3D signal on-the-fly within the MCMC framework \cite{Greig:2015} and \textit{(ii)} emulators \cite{Kern2017,Schmit:2018,Mondal20,Cohen:2020,Ghara20} or deep learning approaches (artificial or convolutional neural networks, \cite[e.g.][]{Shimabukuro:2017b,Doussot:2019,Gillet:2019}) that are specifically trained on a large data-set of simulations to be able to rapidly explore a specific feature or summary statistic (i.e. 21\,cm power spectrum) of the observed 21\,cm signal. These latter methods can in principle enable the more physically accurate but computationally expensive simulations to be explored in a Bayesian context.

The new upper-limits on the 21-cm signal from the SKA pathfinders have reached the point where they are astrophysically interesting. That is, extreme models of the EoR and CD can begin to be disfavoured, while also tangibly highlighting the technological advances that have made with the SKA pathfinders. Using some of the inference frameworks described above we have explored the physical conditions of the IGM, the properties of the high-$z$ galaxies and information about the excess radio background from LOFAR \cite{Ghara20,Greig21,Mondal20}, the MWA \cite{Ghara:2021,Greig:2021b} and HERA \cite{HERA2021b}. They disfavour: \textit{(i)} cold reionisation scenarios driven by low X-ray luminosities of the first galaxies (i.e. whereby the IGM undergoes no heating) with lower limits placed on the IGM spin temperature and \textit{(ii)} extreme radio backgrounds that have been proposed to explain the unexpectedly deep EDGES absorption feature.

\section{Conclusion}
\label{sec:summary}

The future SKA telescope will offer a novel view of the Epoch of Reionization (EoR) and Cosmic Dawn (CD), constraining the processes by which the first stars and galaxies were born and ionized the intergalactic medium.
Currently, SKA-Low pathfinder experiments are hard at work understanding optimal instrument designs and analysis techniques for measuring the 21\,cm signal and mitigating the dominant systematic effects associated with low-frequency intensity mapping.
These pathfinder experiments, the MWA, LOFAR, and HERA, have complementarities that allow for a robust and wide-ranging effort to make a first detection of the signal, and in doing so demonstrate viability of SKA 21\,cm science. In particular, they observe both overlapping and distinct parts of the sky (\autoref{fig:fields}), which can allow for external cross-check measurements of the same field, as well as an understanding of the benefits of other fields where foregrounds may be better behaved.
They also face similar systematic challenges in the shared burden of modeling the foreground and instrumental response to extremely high precision: often studies done with one instrument may reveal particular systematics that are more difficult to understand with another instrument.
For example, the MWA GLEAM point source catalogue \citep{hurley-walker_2017} is a key component of HERA's absolute calibration pipeline, which would have otherwise faced more difficulties in modeling point source foregrounds to such precision.

Shared challenges have allowed for the adoption of analysis algorithms between experiments.
For example, radio frequency interference (RFI) mitigation software developed for LOFAR \citep{Offringa12} is also used on the MWA, and the same ultrafaint RFI mitigation software \citep{wilensky_absolving_2019} is used by the MWA and HERA. Additionally, redundant calibration techniques have been successfully applied to both HERA and the MWA \citep{li_comparing_2018, Dillon2020}, and statistical foreground modeling software has shown promise on both LOFAR and HERA data \citep{Mertens20, Ghosh2020}.

At the same time, these experiments have key differences in their design that make them uniquely situated for specific analyses.
LOFAR, for example, is optimized for high dynamic range imaging, direction-dependent calibration, and point source characterization, which allows it to make precise foreground models needed for its foreground subtraction approach.
HERA, on the other hand, is optimized for maximum sensitivity to a visibility-based 21\,cm power spectrum estimator, has a temporally-stable instrument response, and leverages its high degree of redundancy for calibration.
The MWA strikes a balance between the two with a partial random and redundant layout, and is able to implement both image-based and redundancy-based analyses.

Going forward, the analysis work being done between the pathfinders will promote SKA-era science by answering a few key questions, including how can foregrounds be best modeled, subtracted, or mitigated to enable a 21\,cm measurement? To what precision does the instrumental response need to be understood, and how can this be achieved in real-time for large-antenna arrays? What are the limiting factors in an image-based and a redundancy-based analysis?

Recent developments on these fronts by SKA pathfinders have yielded improved upper limits on the 21\,cm power spectrum at the EoR and CD (\autoref{fig:limits}).
Interpretation of these limits require detailed modeling of the 21\,cm signal, which also face a host of numerical challenges (\autoref{sec:theory}). Nonetheless, these models show that the recent 21\,cm limits are beginning to disfavor extreme astrophysical scenarios of the EoR and CD. Going forward, alternative techniques leveraging higher order statistics and machine learning capabilities are being devised to extract more information from 21\,cm measurements.

While key challenges remain in producing a first, robust 21\,cm measurement, the combined efforts of the SKA pathfinder experiments are exploring a multitude of analysis approaches that are in many ways highly complementary.
These efforts, in addition to the efforts of the wider EoR and CD 21\,cm communities, have in turn made significant progress in recent years in understanding how to enable the transformative science of 21\,cm intensity mapping in the SKA era.

\subsection* {Acknowledgments}
The authors would like to acknowledge the organizers of the 2021 SKA Precursor Conference, where this work was originally shared.
GB acknowledges 
support from the Ministero degli Affari Esteri della Cooperazione Internazionale - Direzione Generale per la Promozione del Sistema Paese Progetto di Grande Rilevanza ZA18GR02. 
Parts of this research were supported by the Australian Research Council Centre of Excellence for All Sky Astrophysics in 3 Dimensions (ASTRO 3D), through project number CE170100013.
NK gratefully acknowledges support from the MIT Pappalardo fellowship.



\bibliography{report}   
\bibliographystyle{spiejour}   







\listoffigures
\listoftables

\end{spacing}
\end{document}